\documentstyle{mn}
\input{epsf}

\begin{document}

\title[Velocity dispersion in NGC 488]
{The shape of the velocity ellipsoid in NGC 488}

\author[J. Gerssen, K. Kuijken and M. R. Merrifield]{Joris Gerssen$^1$,
Konrad Kuijken$^1$ and Michael R. Merrifield$^2$\\
$^1$Kapteyn Institute, Groningen 9700 AV, The Netherlands\\
$^2$Department of Physics and Astronomy, University of Southampton,
Highfield, SO17 1BJ.
}

\maketitle

\begin{abstract}

Theories of stellar orbit diffusion in disk galaxies predict different
rates of increase of the velocity dispersions parallel and perpendicular
to the disk plane, and it is therefore of interest to measure the
different velocity dispersion components in galactic disks of different
types.  We show that it is possible to extract the three components of
the velocity ellipsoid in an intermediate-inclination disk galaxy from
measured line-of-sight velocity dispersions on the major and minor axes. 
On applying the method to observations of the Sb galaxy NGC 488, we find
evidence for a higher ratio of vertical to radial dispersion in NGC 488
than in the solar neighbourhood of the Milky Way (the only other place
where this quantity has ever been measured).  The difference is
qualitatively consistent with the notion that spiral structure has been
relatively less important in the dynamical evolution of the disk of NGC
488 than molecular clouds. 

\end{abstract}


\section{Introduction}

It is observationally well established that the velocity dispersion of
main sequence stars increases with advancing spectral type.  This fact
has been recognised ever since velocity dispersions were first
measured in the solar neighbourhood (e.g.  King 1990).  Initially
these observations were explained as the result of equipartition of
energy because the mass of the stars decreases along the main
sequence.  However the two-body relaxation time scale is much too long
to have any effect on the stellar velocity ellipsoid, prompting
explanations relying on collective effects instead.  Some of the
different functional forms that have been suggested for the velocity
dispersion-age relation are reviewed by Lacey (1991).  This increase
in velocity dispersion, or heating, depends on the roughness of the
gravitational potential in the disk, so knowledge about the shape of
the velocity ellipsoid will tell us a great deal about the dynamical
history of a disk.

Until recently direct measurements of the three-dimensional shape of the
velocity ellipsoid have been restricted to the solar neighbourhood. 
Observations of the stellar velocities in external spiral galaxies
have concentrated on systems that are either close to edge-on or face-on
(van der Kruit and Freeman 1986; Bottema 1995) and will therefore only
provide information about a single component of the velocity dispersion. 
From these measurements only indirect inferences can be drawn about the
shape of velocity ellipsoids in galaxies because the results of a small
sample of {\it different} galaxies are being compared in a statistical
way.  Moreover, this shape will be subject to rather large uncertainties
because the errors of the face-on and edge-on galaxies are compounded,
and because relating face-on and edge-on galaxies is rather delicate.

In this pilot study we show that it is possible to derive the shape of
the velocity ellipsoid within a {\it single} galaxy.  We use the fact
that an intermediate-inclination galaxy shows different projections of
the velocity dispersion at different galactocentric azimuths.  In
section 2 the method by which we extract the velocity dispersions
information is described.  In section 3 we apply this analysis to the
large early type spiral NGC 488 and in section 4 we discuss the
results obtained for NGC 488 and compare them to the velocity
ellipsoid in the solar neighbourhood.

\section{Analysis}

In cylindrical polar coordinates $(R,\phi,z)$ the line-of-sight
velocity dispersion as a function of (intrinsic) position angle
$\phi$ in a thin axisymmetric disk is

\begin{equation}
\left[ \sigma_R^2 \sin^2 \phi +\sigma_{\phi}^2 \cos^2 \phi \right]
\sin^2 i +\sigma_z^2 \cos^2 i ,
\end{equation}
which can be written as

\begin{equation}
\frac{1}{2} \sin^2 i \left[ (\sigma^2_R + \sigma^2_{\phi} + 2 \sigma^2_z 
\cot^2 i) - (\sigma^2_R - \sigma^2_{\phi}) \cos 2\phi \right].
\end{equation}

Thus, it consists of an element with a $\cos 2\phi$ variation that
depends only on the components of the dispersion in the plane of the
galaxy, and an element with no dependence on $\phi$ that depends on
all three components of the velocity dispersion.  Observations along
at least two axes (preferably the major and minor axes, which provide
maximum leverage), are therefore required to extract both
coefficients.  Furthermore, in disk galaxies, in which most orbits are
well-described by the epicycle approximation the radial and azimuthal
dispersions obey the relation

\begin{equation}
\label{eq:ax}
\frac{\sigma^2_{\phi}}{\sigma^2_{R}} = 
{1 \over 2}\left(1+ {\partial \ln V_c \over \partial \ln R}\right),
\end{equation}
where $V_c$ is the circular speed in the galaxy.  (In the solar
neighbourhood, this expression reduces to $-B/(A-B)$, where $A$ and
$B$ are the Oort constants.)  Within the epicycle approximation the
stellar rotation speed $\overline{V}(R)$ equals the circular speed
$\sqrt{R\partial\Psi/\partial R}$, so the righthandside of
eq.~\ref{eq:ax} can be derived from the same spectral observations
that are used to measure the velocity dispersions.  The effect of
higher-order approximations on this formula are discussed by Kuijken
\& Tremaine 1991, who showed that the strongest deviations from
eq.~\ref{eq:ax} are to be found at radii of several disk scale lengths
(see also Cuddeford \& Binney 1994).

Where the circular speed $V_c$ can be measured separately (e.g., from an
emission-line rotation curve), a further constraint on the dispersions
and velocities is the asymmetric drift equation,
\begin{equation}
V_c^2-\overline{V}^2=
\sigma_R^2 \left[ \frac{R}{h}-R\frac{\partial}{\partial R}
\ln(\sigma_R^2)-\frac{1}{2}+\frac{R}{2 V_c} \frac{\partial V_c}
{\partial R} \right] -R\frac{\partial \sigma^2_{R z}}{\partial z}
\end{equation}

The last term of equation 4 describes the tilting of the velocity
ellipsoid.  The two limiting cases of this tilting term are zero and
($\sigma_R^2-\sigma_z^2$). Orbit integration by Binney \& Spergel
(1983) and by Kuijken \& Gilmore (1989) suggest that in the solar
neighbourhood the truth lies close to midway between the two extremes.
As will be seen below, the uncertainty in this term is not a concern
in the present analysis.

In summary, the three components of the stellar velocity ellipsoid can be
deduced from measurements of the line-of-sight dispersions along two
position angles in a galaxy disk. The two sets of measurements,
together with the ratio of tangential and radial velocity dispersions
appropriate for nearly circular orbits, provide the three equations
necessary to deproject the ellipsoid. If the asymmetric drift can also
be measured, e.g. when a rotation curve for cold interstellar gas is
available, the system is overdetermined, allowing a consistency check
on the result.

\section{NGC 488}

We choose the large early type spiral NGC 488 for this analysis
because of its regular optical appearance and its intermediate
inclination.  Table 1 lists some properties of this galaxy.  Note the
very high rotational velocity (Peterson 1980). 

\begin{table}
\caption{Parameters of NGC 488}
\begin{center}
\begin{tabular}{lc}
\hline
Hubble type & Sb \\
Inclination & $40^{\circ}$ \\
Distance & 30 Mpc (for $H_0$ of 75) \\
Max. rotational velocity & 360 km/s \\
Angular size & 5.2 x 3.9 arcmin \\
Photometric scale length & 40" in B \\
\hline
\end{tabular}
\end{center}  
\end{table}

B and I band images of NGC 488 were obtained with the 48 inch telescope
at Mt.\ Hopkins observatory in September 1992.  Bulge-disk
decompositions were performed on these images to assess the extent to
which the bulge contaminated the velocity dispersions of the disk.  The
light profiles obtained from these images along the major and minor axis
are textbook examples of an exponential disk and a $R^{1/4}$ bulge, see
fig.  1.  After subtracting a straight line fit to the linear part of
light profile we found that beyond a radius of 20 arcsec the
disk-to-bulge light ratio increases rapidly.  At 20 arcsec this ratio
is about 7 and at 30 arcsec it is already 25.  Data points at radii
smaller than 20 arcsec were therefore excluded from the analysis.  The
derived scale lengths ($\sim$ 40 arcsec along the B band major axis) are
in good agreement with the literature values (e.g. Shombert \& Bothun 1987). 
The ratio of the minor to major axis scale lengths implies an
inclination of $41^{\circ} \pm 6^{\circ}$, again in good agreement with
the literature value of $40^{\circ}$.  The scale lengths of the I band
image are approximately 20\% shorter than the B band scale lengths.  This
effect is usually attributed to obscuration by dust, but de Jong (1995)
finds that such colour gradients are best explained by differences in
the star formation history as a function of radius. 

\begin{figure}
\epsfxsize=\hsize 
\epsfbox{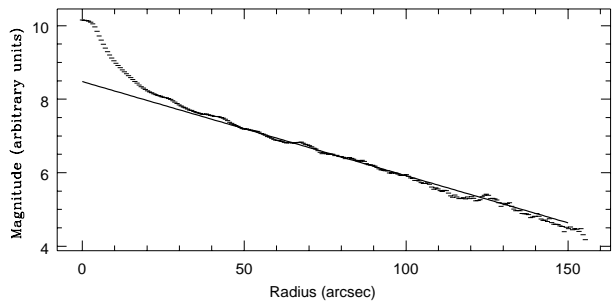}
\caption{Major-axis I-band light profile of NGC 488. The line shows an
exponential fit to the disk luminosity.}
\end{figure}

\subsection{Stellar absorption line data}

Two longslit spectra of NGC 488 were obtained with the Multiple Mirror
Telescope in January 1994 using the Red Channel Spectrograph: a 3 hour
integration spectrum along the major axis and a 2 hour integration
along the minor axis.  Both spectra were centered around the Mg b
triplet at 5200\AA.  Calibrating arc lamp exposures were taken every
30 minutes.  The dispersion per pixel is $\sim$ 0.7\AA.

The longslit spectra were reduced to log-wavelength bins in the standard
way, using {\sc IRAF} packages.  Adjacent spectra were averaged to
obtain a signal-to-noise ratio of at least 25 per bin, and the
absorption-line profiles of these co-added spectra were analysed.

Analysis of the absorption line profiles using the algorithms of
Kuijken \& Merrifield (1993) and van der Marel \& Franx (1993)
revealed no significant departures from a gaussian distribution in the
stellar velocity distributions, so all velocity dispersions were
derived by the traditional gauss-fitting methods using a single
template star (HD 2841, spectral type K5III) to spatially co-added
spectra. We note that the models of Kuijken \& Tremaine (1991), based
on the Shu (1969) distribution function, predict small skewness for
the distribution of azimuthal velocity distribution within 1 disk
scale length; such a marginally non-gaussian shape would not affect our
analysis significantly.

\subsection{Analysis}

\begin{figure*}
\epsfxsize=\hsize
\epsfbox{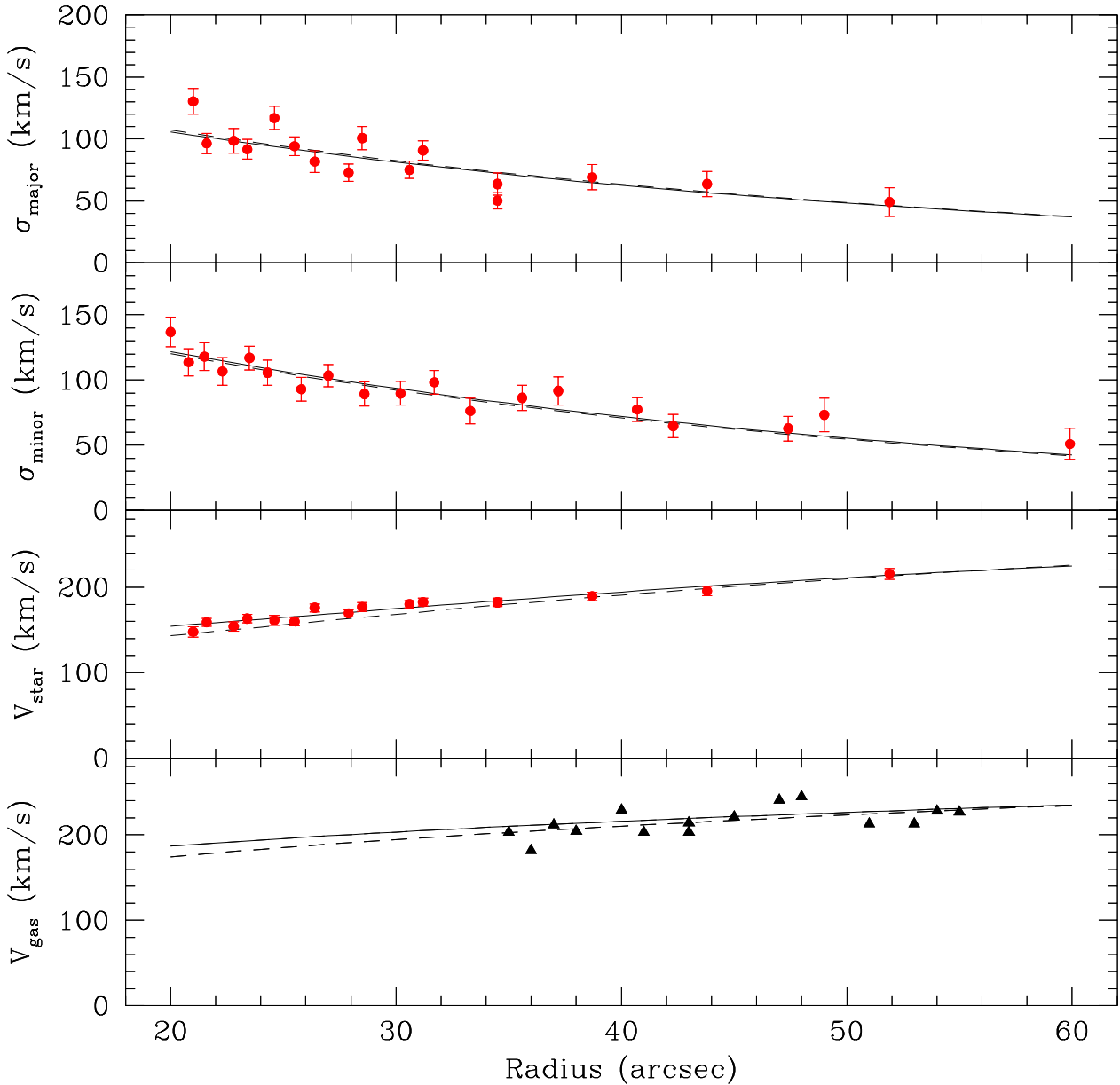}
\caption{The best-fitting model distributions (solid line is fit 1 and
the dashed line is fit 2) and the data. At radii
smaller then 20 arcsec the bulge contaminates the disk light so data
within that region were not included in the fit.
The triangles in the lowest panel are the emission-line data
of Peterson (1980).}
\end{figure*}

We applied the analysis described in section 2 to the spectra.  Due to
the noise present in the data we could not obtain the three components
of the velocity dispersion directly (Merrifield \& Kuijken, 1994) and
adopted therefore a model fitting-approach.  We assumed the following
models for the velocity dispersions: 
\begin{eqnarray}  
\sigma_R&=&\sigma_{R,0} \exp(-R/a);\\
\sigma_z&=&\sigma_{z,0} \exp(-R/a). 
\end{eqnarray}
We also assumed that the circular velocity could be described by a
power-law $V_c=V_{40} (R/40")^{\alpha}$ and used equation 4 to relate
$V_c$ to the stellar rotation speed.

These model distributions were fitted simultaneously to our observables
$\sigma_{\rm minor}$, $\sigma_{\rm major}$ and $\overline{V}(R)$ using
the non linear fit routines described in Press et al.~(1986). 

The best-fit parameters that we obtained are shown in table 2.  In fit
1 the tilting term in equation 4 is set to zero, while in fit 2 it set
equal to the other extreme possible value, $\sigma_R^2-\sigma_z^2$.
Both fit the data equally well.  The $\chi^2$ value of fit 1 is only
marginally better than that of fit 2, and so these data cannot
distinguish between the two extreme models for the tilting term.  It
is, however, interesting to note that sufficiently accurate data might
allow the tilting term to be measured by such a comparison, perhaps
providing constraints on the shape of the potential close to the disk.
The error in the derived axis ratio of the velocity ellipsoid is
slightly larger than one would naively expect because $\sigma_R$ and
$\sigma_z$ are anti-correlated.  Adopting the average between fits 1
and 2 as representing the most realistic tilting term, we obtain
\begin{equation}
\frac{\sigma_z}{\sigma_R}=0.70 \pm 0.19.
\end{equation}

\begin{table}
\caption{Best-fitting parameters for the model distributions. The quoted
errors are one sigma errors.}
\begin{center}
\begin{tabular}{ccc}
\hline
Parameter & fit 1 & fit 2 \\
 && \\
$V_{40}$ (km/s) & 336 $\pm$ 8 & 327 $\pm$ 5\\
$\alpha$ & 0.21 $\pm$ 0.04 & 0.27 $\pm$ 0.04 \\
$\sigma_{R,0}$ (km/s) & 253 $\pm$ 32 & 237 $\pm$ 38 \\
$\sigma_{z,0}$ (km/s) & 164 $\pm$ 27 & 176 $\pm$ 30 \\
$a$ (arcsec) & 38 $\pm$ 4 & 38 $\pm$ 4\\
 && \\
$\sigma_z/\sigma_R$ & 0.65 $\pm$ 0.16 & 0.74 $\pm$ 0.22 \\
\hline
\end{tabular}
\end{center}
\end{table}

We have tested this procedure on a large ($\sim 500$) set of artificial
major and minor axis spectra, created from the template star to resemble
the observed galactic spectra as closely as possible, with a different
poisson noise realisation for each spectrum.  The best-fit parameters
obtained from these spectra scatter with the same dispersion as the
errors obtained from any individual fit to a spectrum. 
Therefore, we conclude that the parameters returned by the fit programme
are reliable.

An obvious extension to our fitting procedure would be to use different
scale lengths for the $R$ and $z$ dispersion components.  Unfortunately
the data are not of sufficient quality to allow for a six parameter fit. 
One of the six eigenvalues that we obtained from diagonalising the
correlation matrix was much larger than the other five, a clear
indication that a six parameter fit is stretching the data a bit too
much. 

Interestingly, the kinematic scale length of the disk appears to be
comparable to the photometric one.  This is not the expectation from
local isothermal approximation for disks, which predicts a scale length
double the photometric one.  The most likely explanation is probably the
fact that we measure the scale length in the B band while the stellar
mass distribution is best traced in the K band, the near infra-red. 
Empirically it is found that scale lengths in the K band are shorter
than the B band up to a factor of about~2 (de Jong 1995, figure 4 and
Peletier et al.  1996).  Alternatively the approximation of a local
isothermal distribution breaks down. 

\subsection{The emission line data}

The fits described above are to the stellar data only.  The lowest
panel in fig.  2 shows the emission line data of Peterson and the
predictions from our two fits.  Both our fits are fully consistent
with these data, and provide extra confirmation of the validity of our
analysis.  (A simple power-law fit to the emission-line data plotted
in fig.  2 gives a power index of 0.18 $\pm$ 0.13.) Unfortunately
there are no measured emission line velocities in the inner part of
the disk.  Peterson gives velocities around 195 km/s near a radius of
10 arcsec which are a little higher than our fits would predict.
However there is no reason why a power-law rotation curve should
persist into those central, bulge dominated regions.

\section{Discussion}

This is the first direct measurement of the vertical-to-radial
velocity dispersion ratio anywhere outside the solar neighbourhood.
Previous determinations have all been indirect, and hence suffer from
large uncertainities.  The method we have described here is quite
straightforward, and could be applied to many systems.

The derived value of $0.70\pm0.19$, which is effectively the average
ratio near one photometric scale length, is somewhat higher than the
solar neighbourhood value of 0.52 $\pm$ 0.03 that Wielen (1977) derived
from the McCormick sample of K and M dwarfs.  However, the error of 0.03
is the purely statistical error, but the scatter between the published
observational estimates of this ratio suggests that the true error may
be larger (see Lacey 1991). 

Any difference between the two ratios is consistent with the findings
about the relative effects of the two dominant heating mechanisms,
molecular clouds and spiral structure, as we now show. 

Heating by molecular clouds was originally proposed by Spitzer and
Schwarzschild (1951).  They proposed that stars in star-cloud
encounters gain kinetic energy at the expense of the clouds because of
the huge masses of the latter.  Subsequent analysis by Lacey (1984)
and $n$-body simulations by Villumsen (1985) showed however that this
mechanism saturates rather quickly (once the stars have sufficient
energy that they spend most of their time outside the cloud layer the
heating rate drops) and could not fully explain the observed heating.

An alternative proposal to heat the disk, due to Barbanis and Woltjer
(1967), explains the heating as the result of stars scattering from
spiral irregularities in the galactic potential.  Carlberg and Sellwood
(1985) have extended this work and showed that this can indeed heat up
the disk.  However this process cannot heat the stars efficiently in the
vertical direction because the vertical oscillation frequency of stars
is much larger than the frequency at which a spiral wave sweeps past the
stars orbiting the disk.  Hence the stars respond almost adiabatically
to this force.

Giant molecular clouds create large spiral wakes, often much larger than
a cloud itself (Julian and Toomre 1966).  This interplay between clouds
and spiral irregularities is not yet completely understood but it is
clear that they are not independent.  Jenkins and Binney (1990) examined
the combined effects of both processes based on Monte Carlo simulations
of the Fokker-Planck equation describing these processes.  They
expressed the relative importance of heating by spiral structure to
heating by clouds by a parameter denoted $\beta$ and calculated the
corresponding ratio of $\sigma_z/\sigma_R$ (see their figure 2).  From
the observed shape of the velocity ellipsoid and velocity dispersion-age
relations they concluded that in the solar neighbourhood the heating of
the disk is dominated by spiral structure ($\beta \sim 90$). 

The mean surface density of the cloud layer near the sun is 1.8
$M_{\odot}$ pc$^{-2}$ (Clemens, Sanders and Scoville 1988).  From the
FCRAO extragalactic CO survey (Young et al.  1995) a mean surface
density for NGC 488 is derived of 3.5 $M_{\odot}$ pc$^{-2}$.  (Young
et al.  find that the CO layer in NGC 488 is best described by a
uniform distribution with a radius of 1.65 arcmin and a CO flux of
${\rm S_{CO}} = 540 \pm 130$ Jy km/s.), higher than in the solar
neighbourhood. NGC 488 is classified as an Sb galaxy.  It has a very
regular tightly wound spiral pattern (the pitch angle is only
5$^\circ$).  It is therefore quite likely that the potential
associated with this spiral pattern is much smoother that that of our
own Galaxy, which has an Sbc Hubble type.  Both these observations
imply that the parameter $\beta$ must be smaller for NGC 488 than for
the solar neighbourhood.  According to the predictions of Jenkins and
Binney a smaller $\beta$ corresponds to a larger $\sigma_z/\sigma_R$
ratio.  

We do indeed find that the $\sigma_z/\sigma_R$ ratio for NGC~488 is
higher than for the Milky Way, however the difference is only one sigma. 
We conclude therefore that the shape of the velocity ellipsoid that we
have determined is qualitatively consistent with the picture sketched by
Jenkins and Binney. 

The technique we have described in this paper would be straightforward
to apply to a larger sample of disk galaxies.  With higher quality
data, it would also be possible to extend the analysis to map out the
radial variation of the velocity ellipsoid shape, a quantity which has
never been observationally constrained.  Both projects would provide
important measurements for comparison with the theoretical treatments
of the heating processes in stellar disks.

\section*{Acknowledgments}
The data presented in this paper were obtained using the Multiple
Mirror Telescope, which is a joint facility of the Smithsonian
Institute and the University of Arizona.  Much of the analysis was
performed using {\sc iraf}, which is distributed by NOAO.
We  thank the referee, Cedric Lacey for his helpful comments.

\end{document}